\documentclass[12pt]{article}
\usepackage{epsfig}
\usepackage[hang,bf,small]{caption}
\DeclareGraphicsExtensions{.eps.gz,.eps,.ps,.ps.gz}
\parskip=\itemsep               

\newcommand{\beq}{\begin{equation}}
\newcommand{\eeq}{\end{equation}}
\newcommand{\beqar}[1]{\begin{eqnarray}\label{#1}}
\newcommand{\eeqar}{\end{eqnarray}}

\newcommand{\si}{\sigma}

\newcommand{\as}{\alpha_S}

\def\eq#1{{Eq.~(\ref{#1})}}

\def\arnps#1#2#3{  {\it Ann. Rev. Nucl. Part. Sci. }{\bf #1} (19#2) #3}
\def\npb#1#2#3{    {\it Nucl. Phys. }{\bf B#1} (19#2) #3}
\def\plb#1#2#3{    {\it Phys. Lett. }{\bf B#1} (19#2) #3}
\def\prd#1#2#3{    {\it Phys. Rev. }{\bf D#1} (19#2) #3}

\def\zpc#1#2#3{    {\it Z. Phys. }{\bf C#1} (19#2) #3}

\relax
\begin{document}
\title{
{\Large \bf   Non-linear evolution and parton  distributions }\\\
{ \Large \bf   at LHC and THERA energies}}
\author{
{\large  M. ~Lublinsky,\thanks{e-mail: mal@techunix.technion.ac.il}~ 
$\mathbf{{}^{a)}}$ \,~ E.~Gotsman,
\thanks{e-mail: gotsman@post.tau.ac.il} ~$\mathbf{{}^{b)}}$ \,~
E.~Levin,\thanks{e-mail: leving@post.tau.ac.il}~ 
$\mathbf{{}^{b)}}$\,\,and\,\,
~ U.~Maor,\thanks{e-mail: maor@post.tau.ac.il} ~$\mathbf{{}^{b)}}$}\\[4.5ex]
{\it ${}^{a)}$  Department of Physics}\\
{\it  Technion -- Israel Institute of   Technology}\\
{\it  Haifa 32000, ISRAEL}\\[4.5ex]
{\it ${}^{b)}$ HEP Department}\\
{\it  School of Physics and Astronomy}\\
{\it Raymond and Beverly Sackler Faculty of Exact Science}\\
{\it Tel Aviv University, Tel Aviv, 69978, ISRAEL}\\[1.5ex]
}

\date{\today}
\maketitle
\thispagestyle{empty}

\begin{abstract} 
 We suggest a new procedure  for extrapolating  the parton
distributions
 from HERA energies 
to higher energies at THERA and  LHC. The procedure suggested
consists of two
steps: first, 
we solve the non-linear evolution
 equation which includes the higher twists contributions, however this
equation is deficient due to the  
low ($log(1/x)$) accuracy of our
 calculations. Second, we introduce a correcting function for which we write a 
DGLAP type linear evolution
 equation. We show that this correcting function is small in the whole kinematic
 region and decreases at low $x$.
The nonlinear evolution equation is solved numerically
 and first estimates for the saturation scale,
 as well as for the value of the gluon density at THERA and LHC energies 
 are  made. 
We show that
 non-linear effects lead to damping of the gluon density by a factor of
 $2 \div 3$  at 
$x \approx 10^{-7}$.

 \end{abstract}
\thispagestyle{empty}
\begin{flushright}
\vspace{-20.5cm}
TAUP-2667-2001\\
\today
\end{flushright}   
\newpage
\setcounter{page}{1}

\section{Introduction}
\setcounter{equation}{0}

The standard perturbative QCD approach to deep inelastic scattering
processes  is based on two main ideas:
\begin{itemize}
\item \quad the DGLAP evolution equations \cite{DGLAP} for the leading twist
parton distributions;
\item \quad  the belief that if we
start QCD evolution from sufficiently high $Q^2  \approx 2 - 4\, GeV^2$  
the higher twist contributions are small. 
\end{itemize}

  The DGLAP evolution has two principal
difficulties. First,  the evolution predicts a steep growth  of parton distributions
 in the region of low $x$ which  contradicts
the unitarity constraints \cite{GLR}. Hence, we can expect large
corrections to the DGLAP evolution equation, at least in the region of low
$x$. The second problem is of a general nature for perturbative QCD.
 It is well
known that any perturbative series is a asymptotic one. Therefore,
the correct estimates for the errors in the pQCD approach is the value of the next order term.
Practically, this means that if we calculated the  leading order gluon density distribution, say
$xG^{LO}$, and the next to leading order one $xG^{NLO}$, then  the correct
estimates for the errors will be $\Delta xG = xG^{NLO} - xG^{LO}$, 
provided  that the very
same initial conditions are used. Such errors are large, especially at low
$x$\footnote{The widely held  opinion that the NLO parton distributions
describe the experimental data better is based on the fact that we can
 change the initial parton distribution drastically and achieve 
 a better description.}.

 It has been shown  over  the last two decades,  that the generally
held
view on 
higher twist contributions is not correct \cite{HT}. Taking
 into account only two terms in Operator Product Expansion for a parton
distribution ($xG$) we have
$$
xG(x,Q^2)\,\,=\,\,xG^{LT}(x,Q^2)\,\,+\,\,\frac{M^2}{Q^2}\,\,xG^{HT}(x,Q^2)\,\,\,+\,\,\,
O(\frac{M^4}{Q^4})\,,
$$
 with the high twist term
$xG^{HT}(x,Q^2) \,\,\propto \,\, \left(\,xG^{LT}(x,Q^2)\,\right)^2$  in the
region of low $x$ \cite{HT}. The leading twist function $xG^{LT}(x,Q^2)$ sharply increases  at 
$x\,\rightarrow\,0$. Hence, we cannot conclude that the higher twist contributions are
small in the whole kinematic region even if they are small for the initial
value of $Q^2 = Q^2_0$. The scale $M$ is introduced  for dimensional reasons.

In this paper,  we suggest a remedy for these two  difficulties.
 First, we propose
  solving a nonlinear evolution equation which takes into account the
most significant higher twist contributions and which specifies  a high
energy 
(low $x$) behavior of the parton densities, which are in accordance
with the
unitarity
constraints. The parton distributions which we obtain are then amended
by adding 
 to the solution of the nonlinear equation $\tilde{N}$,      a
correcting function $\Delta N $ 
 such that
$ N  \,\,=\,\,\tilde{N}\,\, +\,\, \Delta N $. We
then write   a
linear DGLAP-type equation for $\Delta N$. 
 Since $ \Delta N $ in our approach vanishes at
low $x$ we expect  this function to be small and concentrated in the
region of  moderate $x$.  Consequently, this function should be free
from all difficulties inherent in
 the usual scheme solutions of the DGLAP equation. We would like to
mention here that our approach  is very close to  that of
 Kimber, Kwiecinski and Martin \cite{KKM}. However,  our treatment differs 
in the  practical way that  the programme is realized.

An extrapolation of the available parton distribution to the region of lower
$x$ is a  practical problem for LHC energies.  We need to know the
parton
distribution both for  estimates of the background of all interesting
processes at the LHC, such as Higgs production, and for the calculation
of the
cross sections of the rare processes which the LHC is likely to measure. 
In this paper we will show that non-linear evolution will provide a
considerable taming of the parton distribution in the LHC range of
energies,
by a factor of  two - three  in comparison with  predictions based
on
the DGLAP linear evolution. We show here that the non-linear evolution
 considerably diminishes the value of the gluon density  at $x =
10^{-5} \div 10^{-6}$
which is the energy range of the THERA  electron - proton   collider \cite{THERA}.
Therefore, our calculations provide the first estimates for the
possible
collective effects of high parton density QCD  at THERA  based on the correct
non-linear evolution equation.

The paper is organized as follows. In the next section we  formulate our
approach and  write a  nonlinear equation for $\tilde{N}$ and linear
equation for  $\Delta N$.  Section 3 is devoted to a numerical solution of
the
nonlinear equation. The following Section 4 presents some estimates of
 the corrections
induced by the DGLAP kernel.   The linear BFKL equation is discussed
 in  Section 5. In the
Section 6 we summarize our results and mention  our plans for
the future.

\section{A new  approach to DIS}

The DGLAP equation describes the gluon radiation which leads to an
 increase in  the number
of partons. However, when the parton density becomes large,
annihilation
processes become active and they   suppress the gluon radiation and,
so, they tame the rapid increase of the parton densities at a
new saturation scale $Q_s(x)$
\cite{GLR,MUQI,MV}.  The theoretical approach as well as understanding the physics of
parton saturation is  one of the most challenging problem of QCD which
has stimulated the development of new methods 
\cite{GLR,MUQI,MV,SAT,ELTHEORY,BA,KO,ILM},
surprisingly, all approaches lead to the same nonlinear evolution
equation.

\begin{eqnarray}
\label{EQ} 
  \tilde N({\mathbf{x_{01}}},Y;b)&\,=\,&\tilde N({\mathbf{x_{01}}},Y_0;b)\, {\rm exp}\left[-\frac{2
\,C_F\,\as}{\pi} \,\ln\left( \frac{{\mathbf{x^2_{01}}}}{\rho^2}\right)(Y-Y_0)\right ]\,
+\nonumber  \\ & & \frac{C_F\,\as}{\pi^2}\,\int_{Y_0}^Y dy \,  {\rm exp}\left[-\frac{2
\,C_F\,\as}{\pi} \,\ln\left( \frac{{\mathbf{x^2_{01}}}}{\rho^2}\right)(Y-y)\right ]\,\times
\\ \int_{\rho} \,& d^2 {\mathbf{x_{2}}} & 
\frac{{\mathbf{x^2_{01}}}}{{\mathbf{x^2_{02}}}\,
{\mathbf{x^2_{12}}}} \nonumber 
\left(\,2\,\tilde N({\mathbf{x_{02}}},y;{ \mathbf{ b-
\frac{1}{2}
x_{12}}})-\tilde N({\mathbf{x_{02}}},y;{ \mathbf{ b -
\frac{1}{2}
x_{12}}})\tilde N({\mathbf{x_{12}}},y;{ \mathbf{ b- \frac{1}{2}
x_{02}}})\right) \nonumber
\end{eqnarray}
The equation is written for $N(r_{\perp},x; b) = Im
\,a^{el}_{dipole}(r_{\perp},x; b)$ where $a^{el}_{dipole}$ is the amplitude
for elastic scattering for a dipole of size $r_{\perp}$. The total
dipole cross section  is given by
\beq \label{TOTCX}
\sigma_{\rm dipole}(r_{\perp},x) \,\,=\,\,2\,\,\int\,d^2
b\,\,N(r_{\perp},x;b).
\eeq
 The deep inelastic structure function $F_2$  is related to the dipole
cross section
\beq
\label{F2}
F_2(x,Q^2)\,\,\,=\,\,\frac{Q^2}{4\pi}\,\int\,\,d^2 r_{\perp} \int \,d
z\,\,
|\Psi^{\gamma^*}(Q^2; r_{\perp}, z)|^2 \,\,\sigma_{\rm dipole}(r_{\perp},
x)\,\,,
\eeq 
where the QED wave functions $\Psi^{\gamma^*}$ of the virtual photon are well known
\cite{MU94,DOF3,WF}. The meaning of \eq{F2} is simple since it describes the  two
stages of  DIS\cite{GRIB}. The first stage is the decay of a virtual photon into a colorless
dipole ($ q \bar q $ -pair) which is described by wave function
$\Psi^{\gamma^*}$ in \eq{F2}.
 The second stage is the interaction of the dipole
with the target ($\sigma_{\rm dipole}$ in \eq{F2}). This equation is the
simplest manifestation of the fact that the correct degrees of freedom at
high energies in QCD, are color dipoles \cite{MU94}.

In  the equation (\ref{EQ}), the rapidity $Y=-\ln x$ and $Y_0=-\ln x_0$. The
ultraviolet cutoff $\rho$
is needed to regularize the integral, but it does not appear in physical quantities. In the large
$N_c$ limit (number of colors)   $C_F=N_c/2$.

\eq{EQ} has a very simple meaning:  the dipole of size $\mathbf{x_{10}}$ decays
in two dipoles of  sizes $\mathbf{x_{12}}$ and $\mathbf{x_{02}}$  with the decay probability
given by  the wave
function  $| \Psi|^2
\,=\,\frac{\mathbf{x^2_{01}}}{\mathbf{x^2_{02}}\,\mathbf{x^2_{12}}}$. 
 These two dipoles
 then interact with the target. The non-linear term  takes into account
the Glauber corrections for such an interaction. 

The linear part of \eq{EQ} is the BFKL equation\cite{BFKL}, which
describes
the evolution of the multiplicity of the fixed size color dipoles    with
respect to the energy $Y$. The   nonlinear term  corresponds to
a dipole
splitting into two dipoles and it sums the high twist  contributions. 
Note, that the linear part of \eq{EQ} (the BFKL equation)  also has
higher  twist contributions and vise versa, the main contribution of the
non-linear part is to the  leading twist (see Ref.  \cite{MUQI} for
general
arguments and Ref. \cite{HTM} for explicit calculations).

\eq{EQ} was  
suggested in the momentum representation by Gribov, Levin and Ryskin
\cite{GLR}
 and it
was proved in the double log approximation of perturbative QCD by Mueller
and Qiu
\cite{MUQI}, in
Wilson Loop Operator Expansion at high energies by Balitsky \cite{BA}, in
color dipole approach \cite{MU94} to high energy scattering in QCD  by
Kovchegov \cite{KO} and
 in the effective
Lagrangian approach for high parton density QCD by Iancu, Leonidov and McLerran 
(see Ref. \cite{ILM} and
Refs. \cite{ELTHEORY} for previous efforts).
 Therefore, it gives a reliable
tool for an extrapolation of the parton distribution to the region of
low $x$.

One can see that \eq{EQ} does not depend explicitly on the target\footnote{This 
independence is a direct indication that the equation is
  correct for all targets ( hadron and nuclei ) in the regime of high parton density.} and all
such dependence comes from the initial condition at some initial value $x_0$.
For a target nucleus it was argued in Ref. \cite{KO}  that the initial conditions  should be
  taken in the Glauber form:
\beq
\label{ini}
\tilde N(\mathbf{x_{01}},x_0;b)\,=\,N_{GM}(\mathbf{x_{01}},x_0;b)\,,
\eeq
with 
\beq
\label{Glauber}
N_{GM}(\mathbf{x_{01}},x;b)\,=\,1\,\,-\,{\rm exp}\left[ - \frac{\as \pi  \mathbf{x^2_{01}}}
{2\,N_c\, R^2}\,x
G^{DGLAP}(x,  4/\mathbf{x_{01}^2})\,S(\mathbf{b})\right]\,.
\eeq
The equation (\ref{Glauber}) represents the Glauber-Mueller  (GM) formula 
which accounts for the  multiple
dipole-target interaction in the eikonal approximation \cite{DOF3,DOF1,DOF2}.
The function $S(b)$ is a  dipole  profile function inside the target.
The value  of $x_0$ is chosen within the interval 
\beq \label{INCON}
{\rm exp}( - \frac{1}{\as} ) \,\leq \,x_0 \,
\leq \frac{1}{2 m R}\,\,,
\eeq
where $R$ is the radius of the target. In this region the value of $x_0$ is small 
enough to use the low $x$ approximation, but the production of the gluons 
(color dipoles) is still suppressed as $\alpha_S \ln (1/x) 
\,\leq\,\,1$. 
Therefore, in this region we have the instantaneous exchange of the
classical gluon fields.    
Hence, an incoming 
color dipole interacts separately with each  nucleon in a nucleus (see 
Mueller and Kovchegov paper in Ref. \cite{SAT}). 

For the hadron, however, we have no proof that \eq{ini} is correct.  As far
as we understand the only criteria in this problem (at the moment) is the
correct
description of the experimental data. We described almost all available
 HERA data using
\eq{ini} \cite{rep,me},  and we feel confident using \eq{ini} as an 
initial
condition for \eq{EQ}.
In our model we use the Gaussian
$S(b)=e^{-b^2/R^2}$ form for the profile function of the hadron.
 The parameter $R$ is a phenomenological input, 
while the gluon density $xG^{DGLAP}$ is a solution of the DGLAP equation.
 For $x_0$ we have \eq{INCON} for a hadron target as well, but practically 
we choose $x_0 = 10^{-2}$ which satisfies \eq{INCON} and for which we much 
 experimental data  exists to check our initial conditions.

Solutions to the equation (\ref{EQ}) were studied in asymptotic limits in Ref. \cite{LT}. A first
numerical attempt to solve an equation similar to the equation (\ref{EQ})
was reported in Ref. \cite{Braun}. However, 
the solution  was only obtained for nuclei,  and the main emphasize was
put on   extremely
small values of $x$.
In the following section we report on our approach to find 
a numerical solution of the equation (\ref{EQ})  which differs completely from the one adopted in 
\cite{Braun}, as we use the coordinate representation in which
the initial conditions are of a very simple form (see \eq{ini}). The second
reason for using the coordinate representation is the fact that all physical
observables can be expressed in terms of the amplitude for the
dipole-target
interaction in the coordinate  representation (see below for $xG$).

Unfortunately,  equation (\ref{EQ}) is an approximate one. It sums  large $\ln x$ contributions
only. The situation can be improved at small distances, as the exact
$x$ 
dependence of the kernel is known as this is the  DGLAP
kernel.
 An  attempt to obtain
the elastic amplitude $N$  based on  elements of  both the equation 
(\ref{EQ}) and  DGLAP equation was presented in Ref. \cite{KKM}. The authors of that paper
first solve  a generalized DGLAP-BFKL linear equation \cite{KMS}, 
and then add to the solution
a nonlinear perturbation of the form presented in the equation (\ref{EQ}). 
This approach actually
incorporates the  high twist contributions
 in the standard way, treating them as corrections 
to the leading one.

We suggest a different approach to the problem. First,  all twist
contributions should be summed by solving  equation (\ref{EQ}). We
 denote by $\tilde N$  a solution
of  equation  (\ref{EQ}). Second, we  add to the obtained solution a correcting function
$\Delta N$, which will account for the DGLAP kernel:
\beq
\label{Add}
N\,=\,\tilde N\,+\,\Delta N\,.
\eeq
Assuming $\Delta N$ to be small relatively compared to $\tilde N$, we
propose the
 following linear
equation for 
$\Delta N(r_\perp,x;b)=r_\perp^2\,B(2/r_\perp,x;b)$:

\begin{eqnarray}
& &\frac{d B (Q, x;b)}{d (\ln Q^2) }\,=\,\frac{C_F \as}{\pi}
\int_{x/x_0}^1 \,P_{g\rightarrow g}(z)\,\, B(Q, \frac{x}{z};b)\,dz
 -  \label{DN} \\ & &\frac{2\,C_F \as}{\pi}
\int_{x/x_0}^1\frac{d z}{z} \,\tilde N(2/Q, \frac{x}{z};b) 
B(Q, \frac{x}{z};b) + \frac{Q^2C_F \as}{4\pi}\,
\int_{x/x_0}^1 \left(P_{g\rightarrow g}(z)-\frac{2}{z}\right)
\tilde{N}(2/Q, \frac{x}{z};b)\,dz 
\nonumber
\end{eqnarray}  
Here $ P_{g\rightarrow g}(z)$ stands for the usual gluon splitting function:
\beq
P_{g\rightarrow g}(z)\,=\,2\,\left[\frac{1-z}{z}\,+\,\frac{z}{(1-z)_+}\,+\,z\,(1-z)\,+\,
\left(\frac{11}{12}-\frac{n_f}{18}\right)\delta(1-z)\right]\,.
\label{split}
\eeq 
 Equation (\ref{DN}) is a linear equation valid in the 
leading $\ln Q^2$ approximation, with $Q^2=4/r_\perp^2$. The  last
term in the equation 
represents the correction which is due to 
the substitution of the BFKL kernel $1/z$ by the correct DGLAP kernel. The first term 
on the right hand side of the equation  (\ref{DN}) is the 
DGLAP evolution for the correcting function $\Delta N$, while the second
 term is the `nonlinear
``interaction'' of the solutions.
The initial condition $\Delta N(r_{\perp 0},x;b)=N(r_{\perp 0},x;b)-\tilde N(r_{\perp 0},x;b)$ 
is a phenomenological input
at some initial distance $r_{\perp 0}=2/Q_0$ to be specified.
 In the present paper we report on our solution of  equation (\ref{EQ}),
while leaving the question of the determination of
 $\Delta N$  to another paper.

\section{Numerical solution of the nonlinear equation }

In this section we report on the exact numerical solution of the equation (\ref{EQ}) 
with the initial
condition (\ref{ini}). For the transverse hadron size the value $R^2=10\,({\rm GeV}^{-2})$ is
taken,  this corresponds to  the value which is obtained
from ``soft" high energy phenomenology \cite{DL,GLMSOFT} and 
 is in agreement with HERA data on J/$\Psi$ photo-production
\cite{HERAPSI}.  For $xG^{DGLAP}(x,Q^2)$ we use the  GRV'94
 parameterization and 
the leading order solution of the DGLAP evolution equation \cite{GRV}.
The kinematic
region where the solution of (\ref{EQ})
 is found, ranges in $x$ from $10^{-2}$, where the initial conditions 
are set,   to $x$ = $10^{-7}$. The maximal distance is taken to be one
fermi.   
The value of the ultraviolet cutoff $\rho$ is  
$2\times10^{-4}\,({\rm GeV^{-1}})$. The  numerical
solutions obtained are checked to be independent of this choice. 
In all previous studies of equation (\ref{EQ}) the formal dependence on the impact
parameter $b$ was omitted. As our first step we too will neglect the $b$-dependence.
Later we will consider  the   impact parameter dependence.

\subsection{Solution without $b$-dependence.}

In this subsection we assume that both sides of the equation (\ref{EQ}) do not
depend on the impact parameter $b$. In a sense, the assumption
formally corresponds to the
$b=0$ case, though the whole approach is based on  the dipole picture
where the impact
parameter is larger than the dipole sizes.

We propose to solve the equation (\ref{EQ}) by the method of iterations.
 Equation (\ref{EQ}) can be rewritten 
\begin{eqnarray}
\label{EQiter}
\tilde N_{i+1}({\mathbf{x_{01}}},Y) \equiv
\tilde N_{i+1}({\mathbf{x_{01}}},Y;b=0)\,=\, 
 \tilde N_i({\mathbf{x_{01}}},Y_0)\,  e^{-\frac{2
\,C_F\,\as}{\pi} \,\ln\left( \frac{{\mathbf{x^2_{01}}}}{\rho^2}\right)(Y-Y_0) }\,
+  \frac{C_F\,\as}{\pi^2} \times \nonumber \\  \\ 
\int_{Y_0}^Y dy \,   e^{-\frac{2
\,C_F\,\as}{\pi} \,\ln\left( \frac{{\mathbf{x^2_{01}}}}{\rho^2}\right)(Y-y)}\,
 \int_{\rho} \, d^2 {\mathbf{x_{2}}} 
\frac{{\mathbf{x^2_{01}}}}{{\mathbf{x^2_{02}}}\,
{\mathbf{x^2_{12}}}} \nonumber 
\left(\,2\,\tilde N_i({\mathbf{x_{02}}},y)-\tilde N_i({\mathbf{x_{02}}},y)
\tilde N_i({\mathbf{x_{12}}},y)\right) 
\nonumber
\end{eqnarray}
We stop the iterations when the relative error between two iterations 
$| (\tilde N_{i+1}-\tilde N_i)/\tilde N_i|\ll1$
in the kinematic region of the interest\footnote{ 
A formal question of the iteration convergence may certainly arise. 
Namely, when two iterations are sufficiently close  it does
 not formally mean that a limit has been  
reached and even if that the limit exists at all. As a counter-example
one
can consider 
the sum of the  series $1/n$,  which diverges though every step decreases. 
We  have to stress that we completely ignore this question. 
Since a solution to the equation exists
and its asymptotics are known, we are confident about correct convergence
of our 
procedure.}.  
Practically, the iteration is stopped when the maximal 
relative error between  last two iterations is smaller
 than one percent, which is of the same order as our numerical
accuracy.
As a zero iteration the initial condition (\ref{ini}) is taken, 
which is a $x$ independent function. 

\begin{figure}[htbp]
\begin{tabular}{c c}
$x \,= \,10^{-7} $ & $x\,=\,10^{-6} $ \\
 \epsfig{file=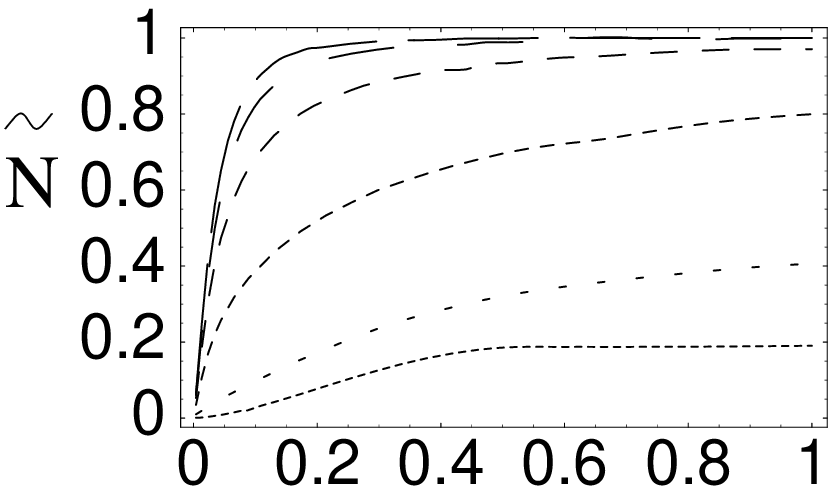,width=70mm, height=40mm}&
\epsfig{file=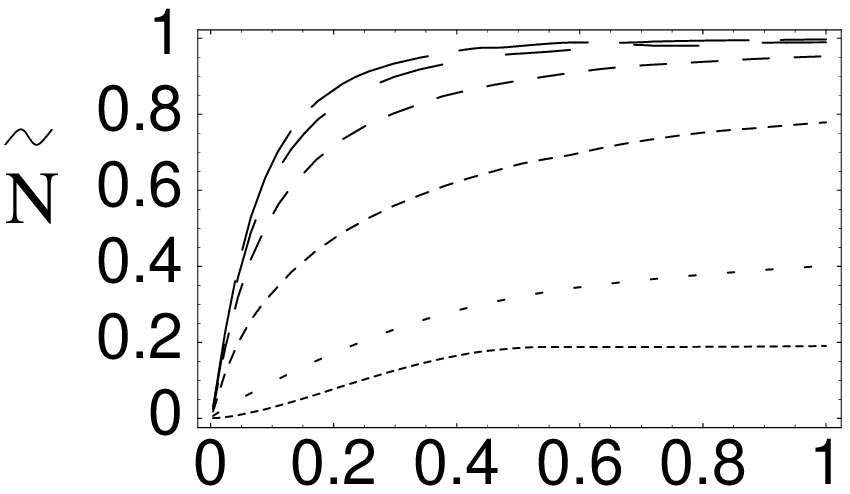,width=70mm, height=40mm}\\ 
$x\,=\,10^{-4} $ & $x\,=\,10^{-3} $ \\
 \epsfig{file=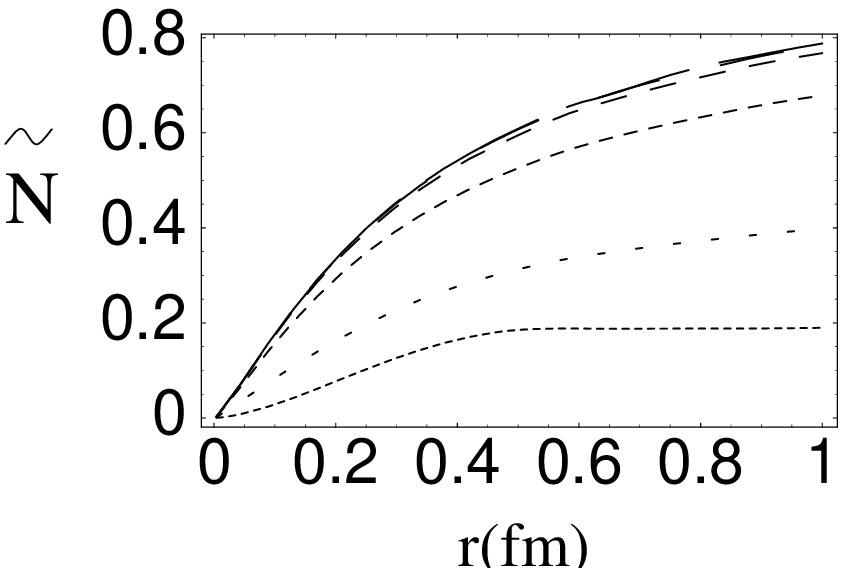,width=70mm, height=50mm}&
\epsfig{file=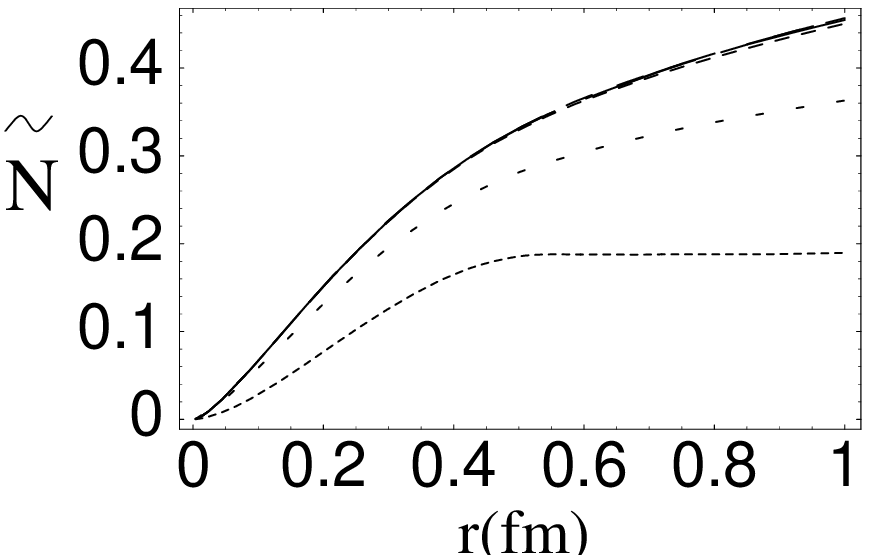,width=68mm, height=48mm}\\ 
\end{tabular}
  \caption[]{\it The function $\tilde N$ is plotted versus distance (in fermi). The six curves
 show the convergence of the iterations (from below 1,5,10,14,18,and 25 iterations). }
\label{fig1}
\end{figure}

Convergence of the solution for the constant value
$\as=0.25$ is shown in the Fig. (\ref{fig1}). As we go  to smaller
values of $x$ the  number
of iterations required for the convergence increases. For $x=10^{-7}$ the solution is obtained
after about 25 iterations. The convergence is very slow because the zero iteration  taken is
 far from the true solution.  Table (\ref{t1}) presents the maximal
relative errors 
for various number of performed iterations.

\begin{table}
\begin{minipage}{9.0 cm} 
\center{
\begin{tabular}{||l||c|c|c|c|c||} 
$x \,\,\backslash$ iter.  &  10  &   14    &    18    &   22    &  25 \\
\hline \hline
 $10^{-7}$ &                    56\% & 24\%  &    9\%  &  5\% & 1\%        \\  
\hline
$10^{-6}$ &                     50\% & 20\%  &   7\%   &  4\%  &  1\%      \\
\hline
 $10^{-5}$ &                    34\% & 10\%  &   3\%   &  1\%   &   1\%    \\
\hline
  $10^{-4}$ &                   14\% & 2\%     &  1\%    &   1\%   &    1\%        \\
\hline
 $10^{-3}$  &                    1\% &  1\%   &    1\%    &  1\%   &    1\%    \\
\end{tabular}}
 \end{minipage}
\begin{minipage}{7.0 cm}  
 \caption{Convergence of the iterations for the $x$ independent zero iteration. 
The table shows the maximal relative errors in percent.}
\label{t1}
\end{minipage} 
\end{table}

\begin{table}
\begin{minipage}{9.0 cm}
\center{
\begin{tabular}{||l||c|c|c|c|c||} 
$x \,\,\backslash$ iter.  &  8  &   9    &    10    &   11    &  12   \\
\hline \hline
 $10^{-7}$ &                    12\% & 9\%  &    6\%  &  3\% & 1\%        \\  
\hline
$10^{-6}$ &                     10\% & 7\%  &    4\%   &  2\%  &  1\%      \\
\hline
 $10^{-5}$ &                    6\% &   4\%  &   3\%   &  1\%   &   1\%    \\
\hline
  $10^{-4}$ &                   6\% &   4\%  &     2\%    &   1\%   &    1\%        \\
\hline
 $10^{-3}$  &                    3\% &  2\%  &      1\%   &  1\%   &    1\%    \\
\end{tabular}}
\end{minipage}
\begin{minipage}{7.0 cm}
\caption{Convergence of the iterations for the GM input for zero iteration. 
The table shows the maximal relative errors in percent.}
\label{t2}
\end{minipage}
\end{table}

In fact, the number of iterations  can be dramatically decreased
 by guessing a better zero iteration.
  The GM formula  (\ref{Glauber}) provides a reasonable choice.
This 
formula happens to be a good approximation to the solution, thus
 significantly reducing
the process of  convergence. Taking the GM formula as the zero iteration, the 
same solution is obtained again after  about a dozen  iterations. 
 Table (\ref{t2}) shows the iteration  convergence.

 Equation (\ref{EQ}) as well as the BFKL equation is derived for a
constant strong
coupling $\as$.  Though some attempts were made to introduce a running coupling  
constant into 
the BFKL equation    it has not yet been done  in a self consistent
way. However,
the use of the running coupling constant in  equation (\ref{EQ}) is
justified in the double
logarithmic  approximation. Motivated by this fact, a solution  of
 equation (\ref{EQ})
with running $\as$ is also obtained.

\begin{figure}[htbp]
\begin{tabular}{c c}
$x \,= \,10^{-7} $ & $x\,=\,10^{-6} $ \\
 \epsfig{file=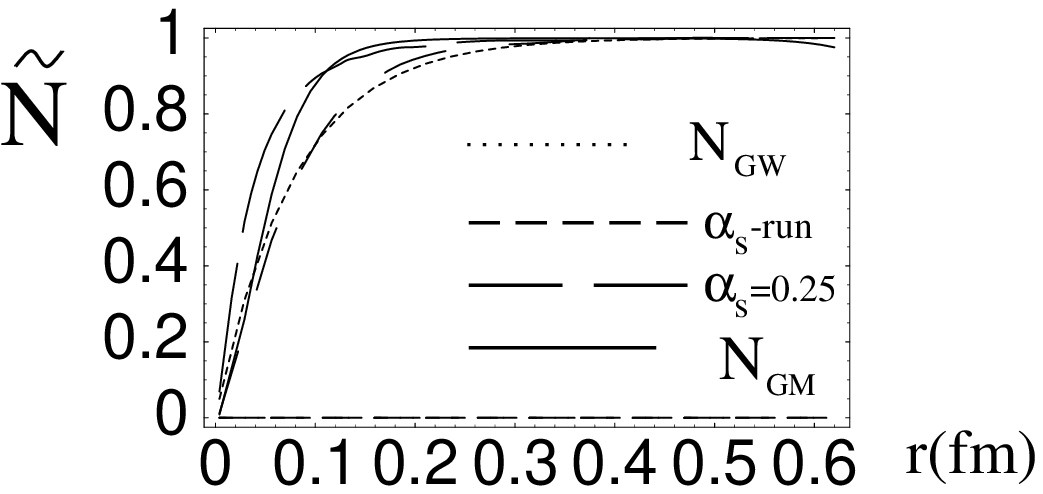,width=82mm, height=40mm}&
\epsfig{file=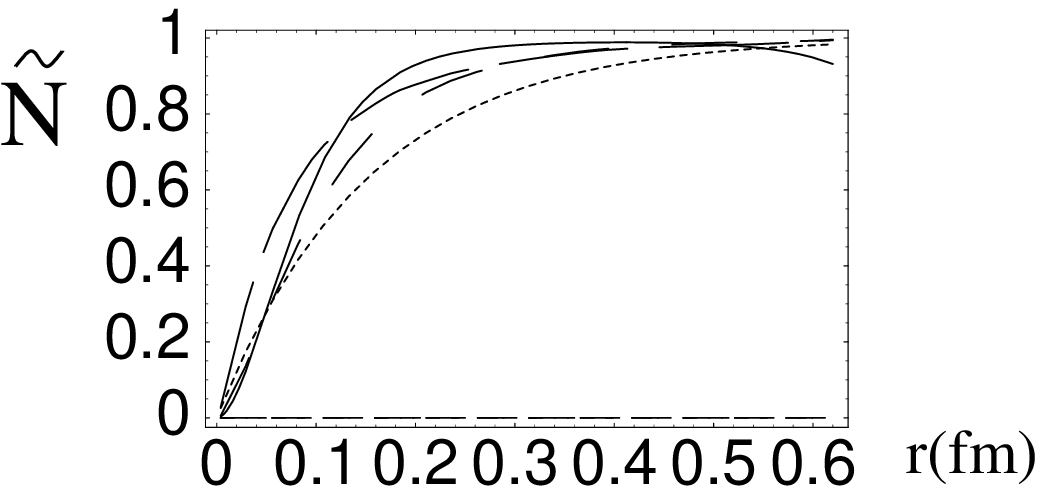,width=82mm, height=40mm}\\ 
$x\,=\,10^{-4} $ & $x\,=\,10^{-3} $ \\
 \epsfig{file=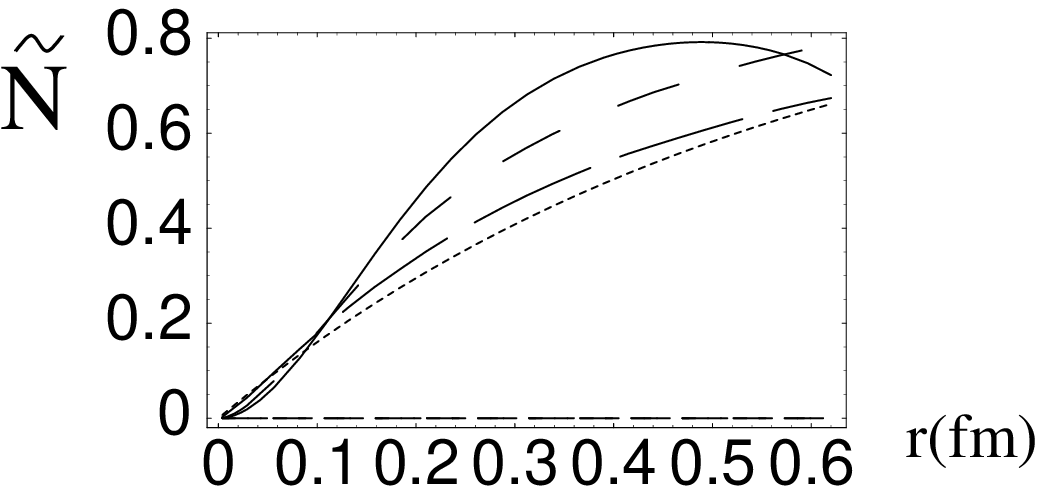,width=82mm, height=40mm}&
\epsfig{file=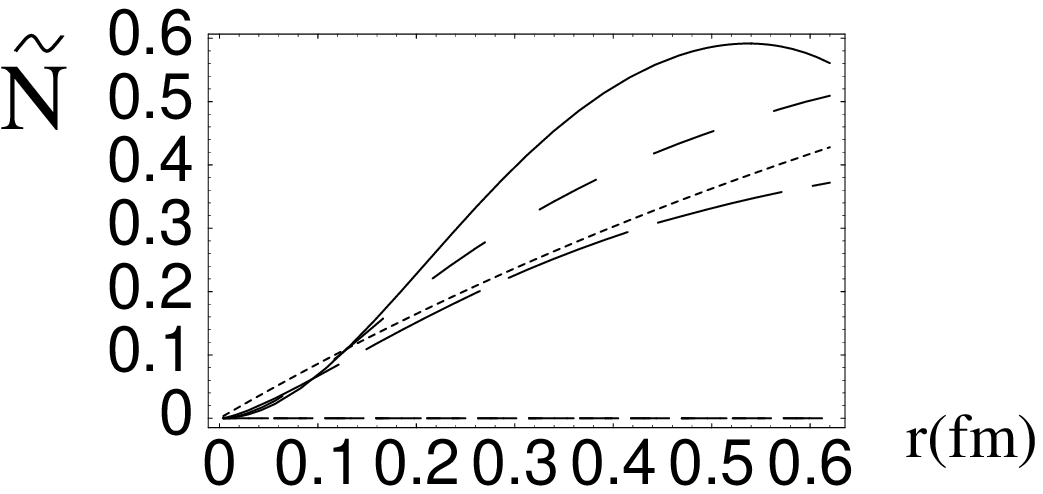,width=82mm, height=40mm}\\ 
\end{tabular}
  \caption[]{\it The comparison between the solutions $\tilde N$,  
Glauber-Mueller formula, and GW model. The four curves correspond to two different
solutions $\tilde N_{\as=0.25}$ (large dashes), $\tilde N_{\as - running}$(small dashes), 
$N_{GM}$ (continuous line), and $N_{GW}$ of the GW model (dots).}
\label{fig2}
\end{figure}

It is interesting
 to compare the  solutions found with two of the widely used saturation
models. 
The first one is the GM  formula (\ref{Glauber}). The second one is the 
Golec-Biernat and Wusthoff (GW) saturation model \cite{WG}.
 In that model  the effective dipole cross section
$\sigma_{\rm dipole}^{GW}(x,r_\perp)$
describing the interaction of the $q\bar q$ pair with a nucleon has the form:
\begin{eqnarray} 
\label{GW} 
\sigma_{\rm dipole}^{GW}(x,r_{\perp})\,=\,\si_0\,\,N_{GW}(x,r_{\perp})\,; \,\,\,\,\,\,\,\,\,
\,\,\,\,\,\,\,
N_{GW}(x,r_{\perp})\,=\,[1\,-\,\exp (-r^2_{\perp}/(4\,R_0^2(x)))]\,;
\\ R_0(x)\,=
\, (x/x_0)^{\lambda/2}
\,({\rm GeV^{-1}})\,;\,\,\,\,\,
 \si_0\,=\,23.03\, ({\rm mb});\,\,\,\, \,\,\lambda\,=\,0.288\,;\,\,\,\,\,\,
\, x_0\,=\,3.04\,\times \,10^{-4}\,.
\nonumber
\end{eqnarray}
The cross section $\si_0$ is a dimension-full parameter which properly 
normalizes the dipole cross
section. A natural comparison is between our function $\tilde N$ and the dimensionless 
saturation function $N_{GW}$ of the GW  model. The Fig. (\ref{fig2})
shows the comparison between our solutions, GM formula, and GW model. It can be seen that
the correct numerical solution is obtained between the GM formula and
GW model.
At large distances  GW $\le \tilde N \le $  GM, while at small distances the inverse situation is 
realized GM $\le \tilde N\le$  GW.  We would like to remind the reader 
that by construction the GM
formula is  a DGLAP solution at small distances.

We would like to add a comment on the behavior of the  GM formula at large
distances.  At fixed $x$ this function has a maximum and then decreases. Such a behavior
is certainly unphysical and  is just an artifact of  using the  GRV parameterization which
itself possesses a similar behavior.

One of the goals of the present research is to determine  the saturation
scale $Q_s(x)$. 
Despite many attempts, no exact mathematical definition of $Q_s(x)$
has so far  has been found. 
However,
some  reasonable estimates can be obtained from the function $\tilde N$.
 As can be seen from the
figures (\ref{fig1},\ref{fig2}) $\tilde N$ behaves in a step like manner as a function of distance: 
at small distances
it tends to zero, while at large distances the saturation value one is approached. 
For such a
 step-like kind of function it is natural to define the saturation
scale 
as a position where $\tilde N=1/2$:
\beq
\label{scale}
\tilde N(2/Q_s, x)\,=\,1/2\,.
\eeq
The equation ({\ref{scale}) defines the saturation scale $Q_s$ as a function of $x$ (Fig.
(\ref{scaleplot},a)).  The saturation scale of the GW model defined as $1/R_0(x)$ is also 
displayed in the Fig. (\ref{scaleplot},a).

\begin{figure}[htbp]
\center{
\begin{tabular}{c c}
\epsfig{file=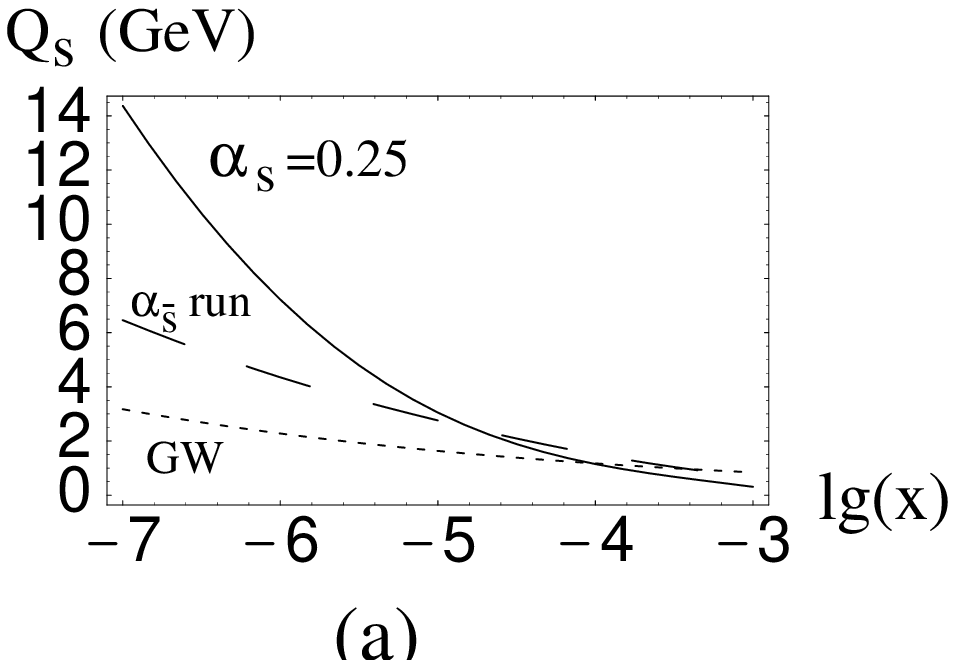,width=70mm, height=50mm} & 
\epsfig{file=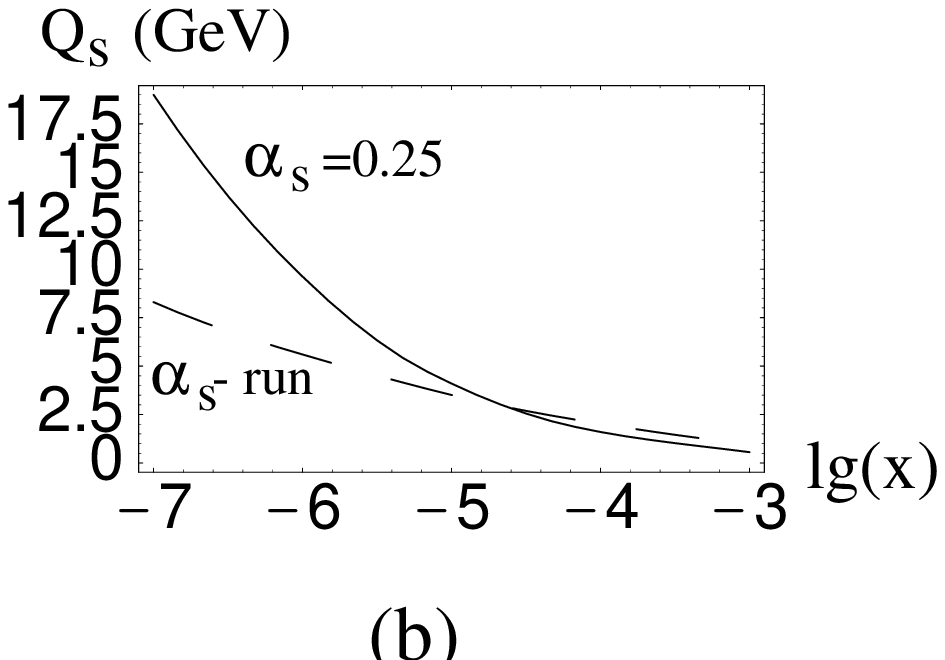,width=70mm, height=50mm}\\
\end{tabular}
}
\caption[]{\it The saturation scale $Q_s$  is plotted as a function $lg x=log_{10}(x)$. 
(a) - the scale obtained from the equation (\ref{scale}), the  dotted line ($GW$) corresponds
to the saturation scale of the GW model; (b)  - the equation (\ref{scale2}) is
used to determine the scale. }
\label{scaleplot}
\end{figure}
We have to comment that the accepted definition of the saturation scale (\ref{scale}) is not
quite consistent with the estimates obtained from the GM formula. In the latter approach 
$Q_s(x)$ is deduced from the requirement that the gluon packing factor in a cascade  
 should be equal to unity. For large values of $x$ ($10^{-2}$ - $10^{-4}$) 
the saturation scale thus obtained is somewhat 
larger than that shown in the Fig. (\ref{scaleplot},a).

\subsection{$b$-dependence of the solution.}

In this subsection we deal with the $b$-dependence of the equation (\ref{EQ}). The proposed
iteration  method can in fact be applied, but in this case the computational time increases
dramatically. Instead, we assume that solution of the equation (\ref{EQ}) preserves 
the very same
$b$-dependence as introduced by the initial conditions (\ref{ini}):
\beq
\label{Nb}
\tilde N(r_\perp,x; b)\,=\, (1\,-\,e^{-\kappa(x,r_\perp)\, S(b)})\,,
\eeq
where $\kappa$ is related to the $b=0$ solution
\beq
\label{kappa}
\kappa(x,r_\perp)\,=\,-\,\ln(1\,-\,\tilde N(r_\perp,x,b=0)).
\eeq
Of course, the assumption requires  verification. In order to verify that (\ref{Nb}) is a solution
of (\ref{EQ}) we should check if it indeed satisfies (\ref{EQ}). So, we
 plug 
$\tilde N(\ref{Nb})$ into the rhs
of (\ref{EQ}) and then compare it with $\tilde N(lhs(\ref{EQ}))$. Fig. (\ref{bcheck})
 shows the comparison 
as a function of  the impact parameter $b$. A  good fit  is  found. The matching
becomes  worse in the region close to  saturation. 
This happens mainly due to problematic numerical  behavior of $\kappa$. 

\begin{figure}[htbp]
\begin{tabular}{c c}
 \epsfig{file=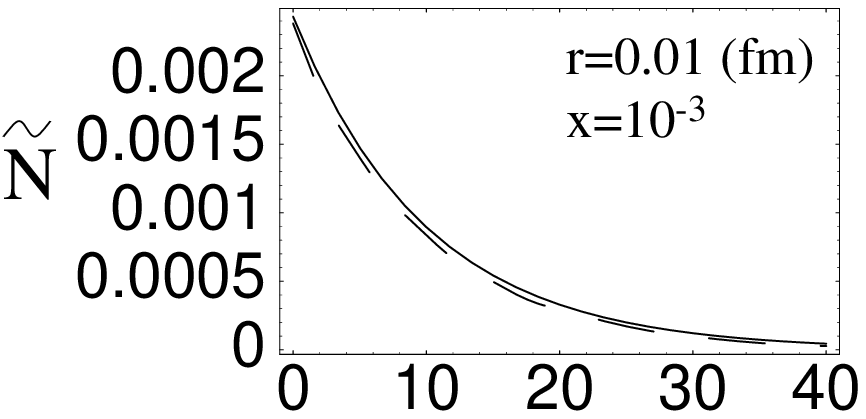,width=75mm, height=42mm}&
\epsfig{file=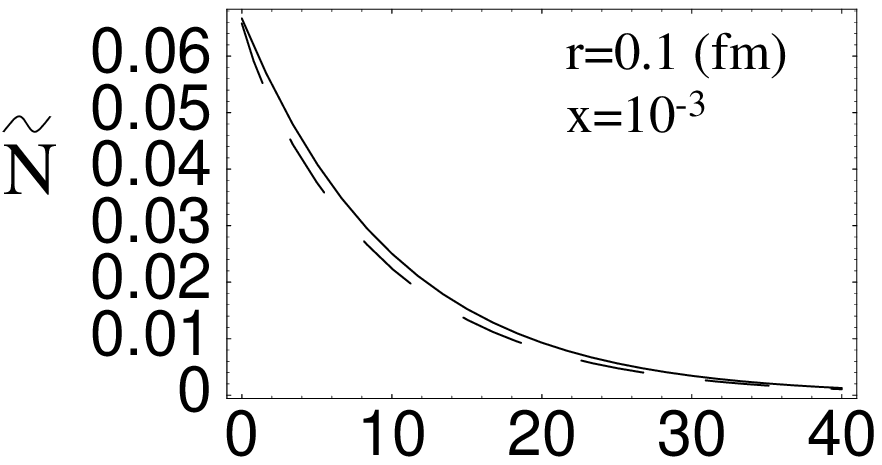,width=70mm, height=40mm}\\ 
 \epsfig{file=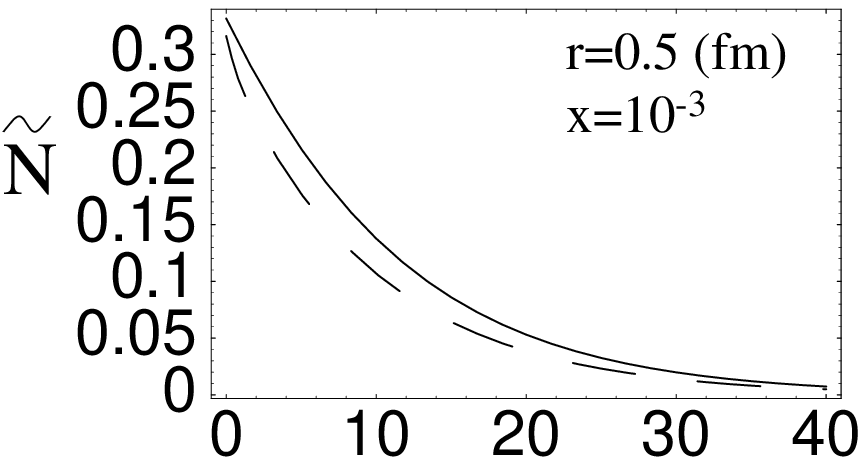,width=70mm, height=40mm}&
\epsfig{file=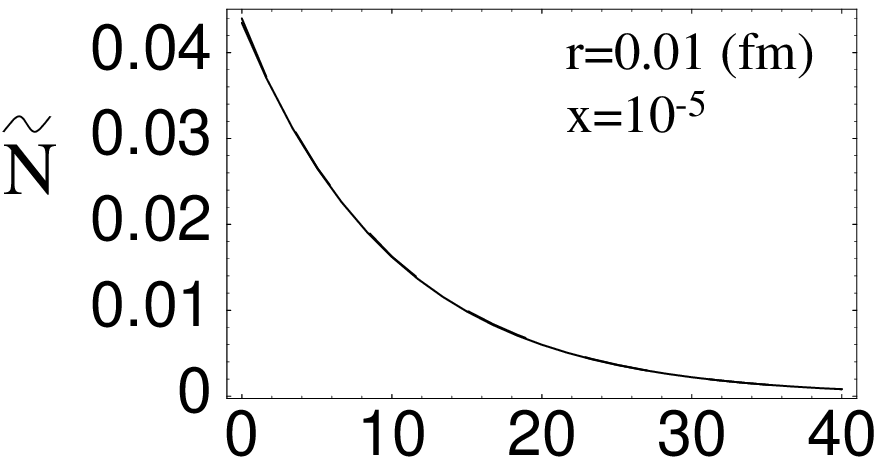,width=70mm, height=40mm}\\ 
 \epsfig{file=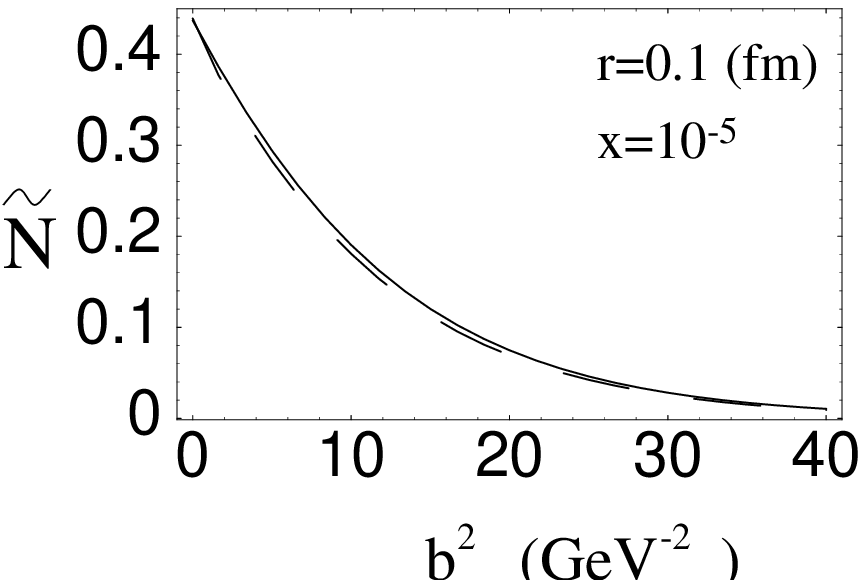,width=70mm, height=45mm}&
\epsfig{file=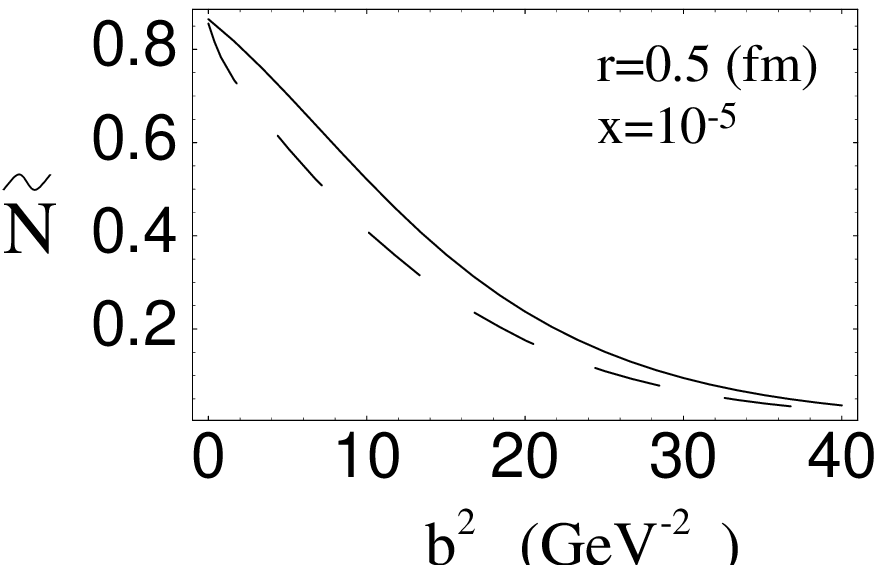,width=70mm, height=45mm}\\ 
\end{tabular}
\caption[]{\it The $b$-dependence of the solution is compared with the model dependence 
of the equation (\ref{Nb}).The graphs are plotted as functions of $b^2$ for two values  of $x$: 
$x=10^{-3}$ and $x=10^{-5}$. The continuous line is the anzatz (\ref{Nb}), 
while the dashed line is 
$\tilde N(lhs(\ref{EQ}))$.}
\label{bcheck}
\end{figure}

In order to make an additional estimate of the accuracy of the anzatz (\ref{Nb}), a solution
of the equation (\ref{EQ}) is obtained for a single value of the impact parameter. 
The only simplification which is made in this computation
is the assumption that the impact parameter $b$ is much larger than the dipole sizes:
\beq
\label{assum}
{\mathbf{x_{01}}}\,\ll\,b; \,\,\,\,\,\,\,\,\,\,\,\,\,\,\,\,\,\,{\mathbf{x_{02}}}\,\ll\,b.
\eeq
In this case the $b$-dependence of the rhs of (\ref{EQ}) is significantly simplified. This is 
 actually the very same approximation that was  used in the previous
subsection. Figure
(\ref{bcheck1}) presents the comparison of the anzatz (\ref{Nb}) with numerically computed
solution for $b^2= 10\,{\rm GeV^{-2}}$.

\begin{figure}[htbp]
\begin{tabular}{c c}
 \epsfig{file=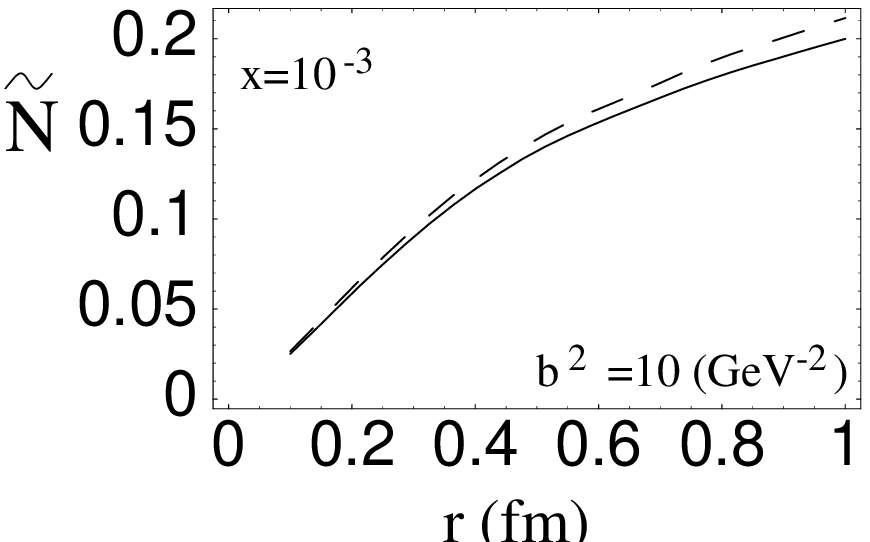,width=75mm, height=42mm}&
\epsfig{file=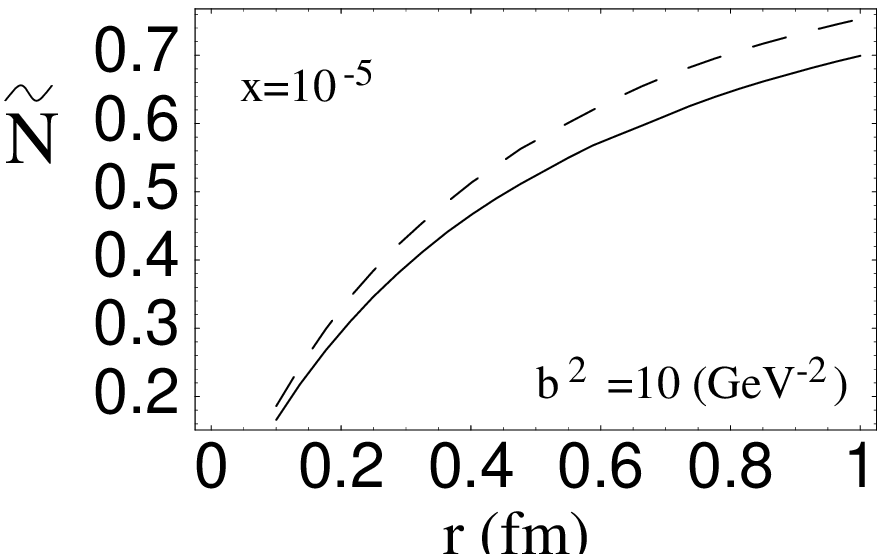,width=70mm, height=42mm}\\ 
\end{tabular}
\caption[]{\it The  solution of the eq. (\ref{EQ}) for $b^2= 10 \,{\rm GeV^{-2}}$
 is compared with the anzatz dependence 
of the equation (\ref{Nb}). The graphs are plotted as functions of distance
 for two values  of $x$: 
$x=10^{-3}$ and $x=10^{-5}$. The continuous
 line is the anzatz  (\ref{Nb}), while the dashed line is the
exact numeric solution.}
\label{bcheck1}
\end{figure}

The match is quite good at moderate $x$, though it becomes worse at smaller
 $x$.
Note, however, from figures (\ref{bcheck}) and
(\ref{bcheck1}) 
we see that the curve of  the anzatz   (\ref{Nb}) is actually 
squeezed by the two curves it was compared to. This provides an
 indication of the overall uncertainty
of the approximation, which we would roughly estimate not to exceed
10\%-20\%.

The above procedure cannot be  claimed to be a proof of the $b$-factorization. 
Moreover, an analytic derivation of the equation (\ref{Nb}) does not seem to be feasible. 
However,
the solution in the form (\ref{Nb}) is clearly a  satisfactory 
approximation of the true solution of  the equation (\ref{EQ}). 

We can proceed now with the evaluation of the dipole cross section
\beq
\label{sidipole}
\sigma_{\rm dipole}(x,r_\perp)\,=\,2\,\int\,d^2b\,\tilde N(r_\perp,x,b)\,.
\eeq
Having assumed (\ref{Nb}), the dipole cross section has the form 
\beq
\label{sidipol1}
\sigma_{\rm dipole}\,=\,2\,\pi\,R^2\,\left[ \ln(\kappa)\,+\,E_1(\kappa)+\gamma\right]\,.
\eeq
In  equation (\ref{sidipol1}) $\gamma$ denotes for the Euler constant,
while $E_1$ is the
exponential integral function. 
The expression (\ref{sidipol1}) predicts the $\ln{\kappa}$ growth of the dipole cross section, 
which is in agreement with the conclusions presented in Ref. \cite{LT}. 
Fig. (\ref{sigratio}) presents a comparison between $\sigma_{\rm dipole}$ from (\ref{sidipol1}) and
the GW model (\ref{GW}).
\begin{figure}[htbp]
\begin{tabular}{c c}
 \epsfig{file=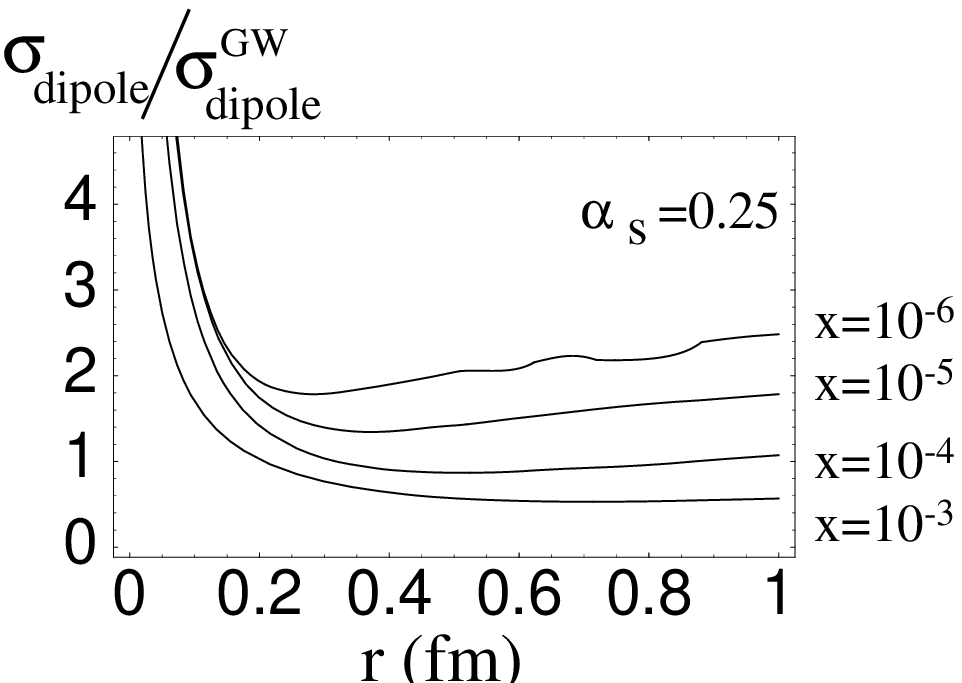,width=70mm, height=50mm} &
 \epsfig{file=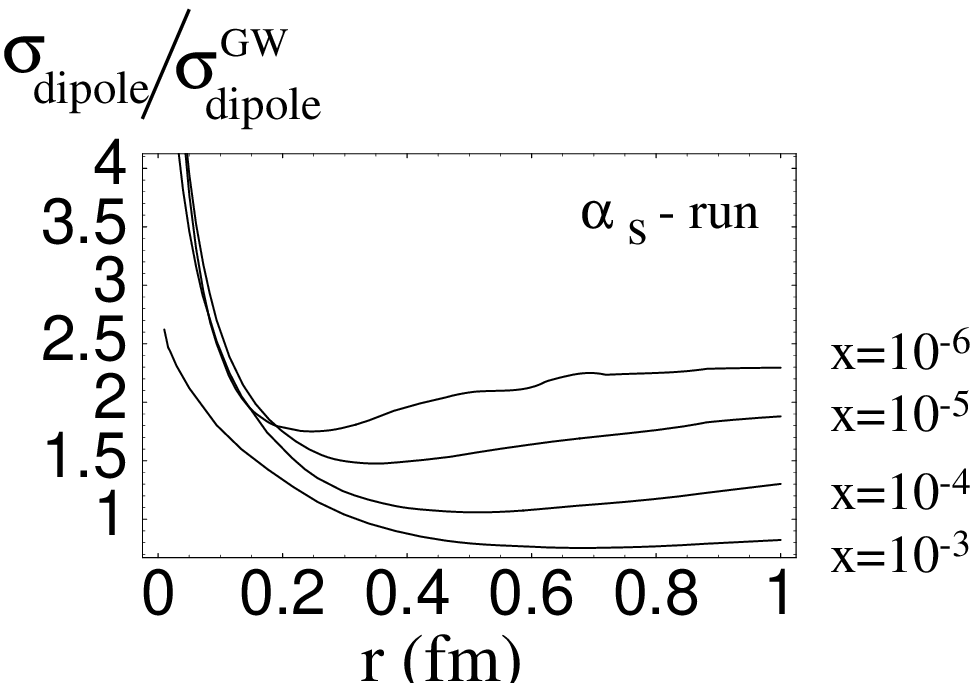,width=70mm, height=50mm} \\
\end{tabular}
\caption[]{\it The ratio $\sigma_{\rm dipole}/\sigma_{\rm dipole}^{GW}$ 
is plotted as a function of 
distance for different values of $x$. }
\label{sigratio}
\end{figure}
 From  Fig. (\ref{sigratio}) it can be seen
 that the ratio $\sigma_{\rm dipole}/\sigma_{\rm dipole}^{GW}$
grows with decreasing $x$. This is in agreement with the  logarithmic growth of the dipole
cross section $\sigma_{\rm dipole}$, while the $\sigma_{\rm dipole}^{GW}$ is 
saturated at small
values of $x$. 

We define the gluon density $xG$  by using the Mueller formula. It relates the gluon density
to the dipole cross section:
\beq
\label{xG}
xG(x,  Q^2)\,=\,\frac{4}{\pi^3} \int_x^1 \frac{dx^\prime}{x^\prime}
\int_{4/Q^2}^\infty\frac{dr^2}{r^4} \,\si_{\rm dipole}(x^\prime,r)\,.
\eeq
Fig. (\ref{xg}) presents the comparison between the gluon density obtained by (\ref{xG})
and the DGLAP input value in the GRV parameterization. At small $x$ DGLAP predicts
steep growth of the gluon density (power law). As was expected, the  shadowing corrections
suppress the growth, and the $x$ growth of the function $xG$ is  logarithmic which does not 
 contradict  unitarity. 
\begin{figure}[htbp]
\begin{tabular}{c c}
\epsfig{file=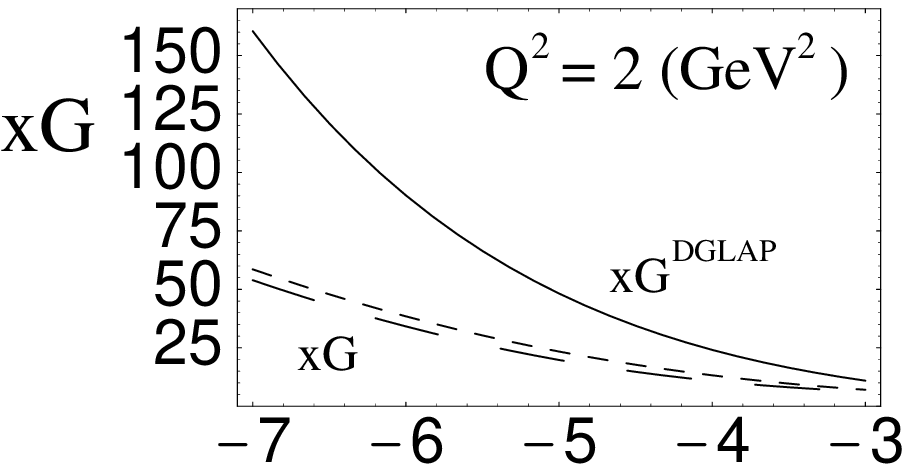,width=70mm, height=40mm}&
\epsfig{file=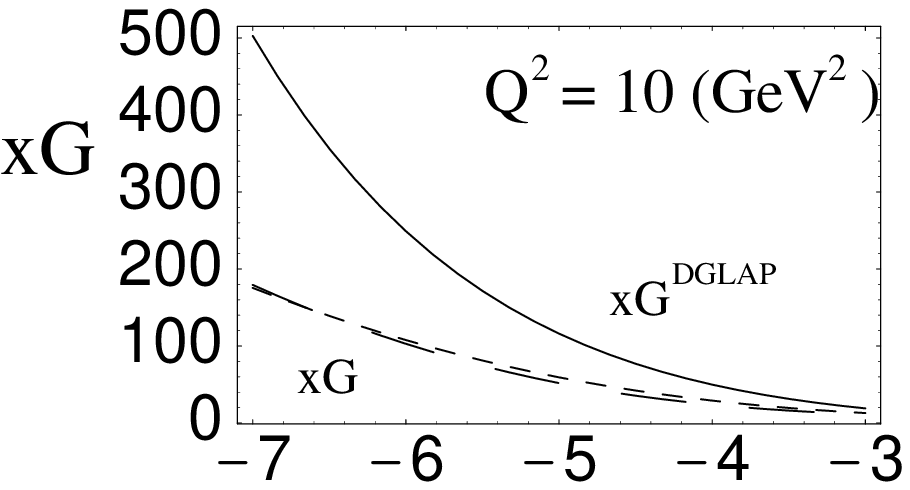,width=70mm, height=40mm}\\ 
 \epsfig{file=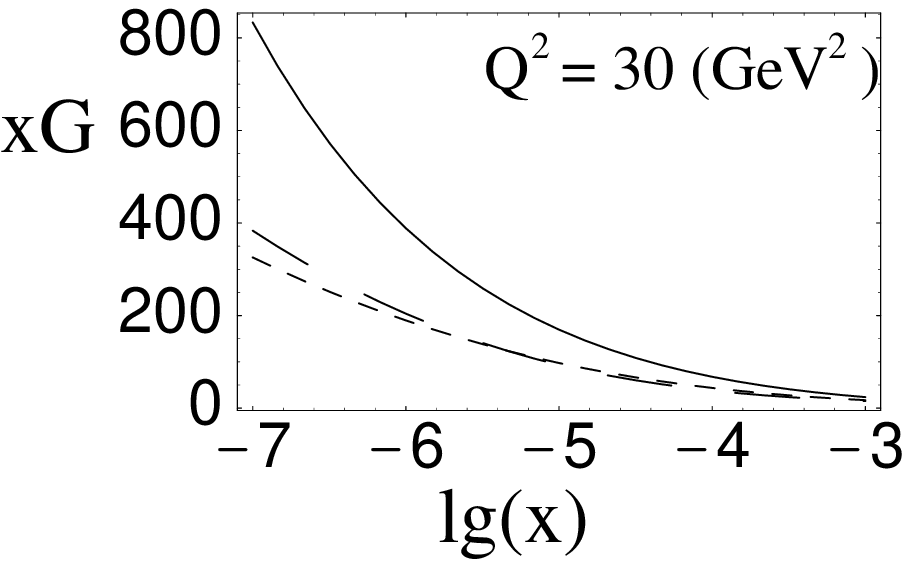,width=70mm, height=50mm}&
\epsfig{file=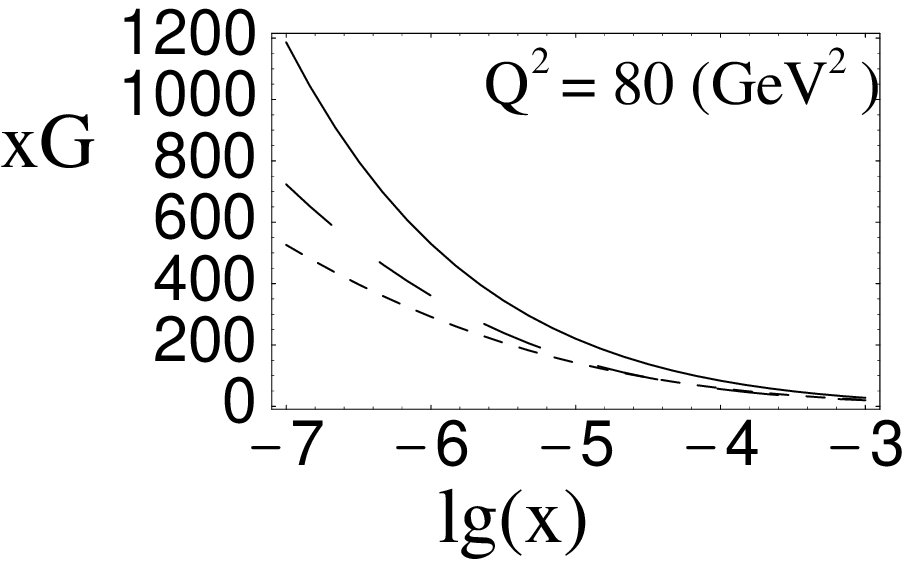,width=70mm, height=50mm}\\ 
\end{tabular}
  \caption[]{\it The function $xG$ is plotted versus $\lg (x)$. The small dashes
  correspond to the solution with the running $\as$, while the large dashes 
are used for  the constant $\as=0.25$. The continuous line is the GRV
parameterization.}
\label{xg}
\end{figure}

The gluon density  $xG$ defined as in (\ref{xG}) 
grows with decreasing $x$. In a sense,
 this   observation contradicts the ``super-saturation'' of Ref. \cite{Braun} where the density is 
predicted to vanish in the large rapidity limit. However, the gluon density definition accepted
in Ref. \cite{Braun} differs from ours, and hence the functions are not compatible.

It is worth  making an additional comment about the saturation scale $Q_s(x)$.
 In the previous
subsection we defined it through the equation (\ref{scale}). Here we  propose an alternative 
definition similar to the one adopted in the GM formula. Namely:
\beq
\label{scale2}
2\,\kappa(x,2/Q_s)\,=\,1\,.
\eeq
The above definition corresponds to $\tilde N(2/Q_s,x,b=0)\simeq 0.4$, and it predicts
a somewhat larger saturation scale $Q_s(x)$ (Fig. (\ref{scaleplot},b))
than the equation  (\ref{scale}). Comparing the four curves of 
both figures  (\ref{scaleplot},a) and  
(\ref{scaleplot},b) an uncertainty in the $Q_s(x)$ determination can be estimated. It is about
30\%-40\% through a very large range of $x$, though reaching 100\%
uncertainty at $x=10^{-7}$.

\section{DGLAP correction - consistency check}

In this section we would like to make some comments regarding the
 consistency of our
approach. It was argued previously that it is necessary to add a
correction term   $\Delta N$    
  to the solution $\tilde N$  
of the nonlinear equation (\ref{EQ}) which we found. 
In  turn, the function $\Delta N$ is a solution of the evolution equation (\ref{DN}). 

Consistency of the approach requires  the support of $\Delta N$ only
 at moderate $x$.
 The function  $\Delta N$ should  give vanishing contributions at very small $x$. 
We also expect this function to decrease with $Q^2$.  In order to check the 
above  conditions some asymptotic estimates can be made
 without explicitly solving the
equation  (\ref{DN}). 

At very small $x$ and large distances the function $\tilde N\simeq 1$.
 For this case   equation 
   (\ref{DN}) can be rewritten:
\beq
\frac{d B (Q, x)}{d (\ln Q^2) }\,=\,\frac{C_F \as}{\pi}
\int_{x/x_0}^1 \,\left(P_{g\rightarrow g}(z)-\frac{2}{z}\right)\, 
\left(\frac{Q^2}{4}\tilde N (2/Q, \frac{x}{z})\,+\,
B(Q, \frac{x}{z})\right)\,dz \,.
\label{DN1}
\eeq
The main observation is that the evolution kernel entering the equation (\ref{DN1}) is actually
negative. Hence the function $\Delta N$ is a decreasing function of $Q$. 

We illustrate the point in a model where the anomalous dimension
has the form \cite{EKL}
\beq
\label{andim}
\gamma(\omega)=\frac{\as N_c}{\pi}\left(\frac{1}{\omega}\,-\,1\right)\,,
\eeq
where the anomalous dimension is defined by the Mellin transform of
the splitting function
$P_{g\rightarrow g}$:
\beq
\label{mellin}
\gamma(\omega)\,=\,\frac{\as\,C_F}{\pi}\int_0^1\,dz\,P_{g\rightarrow g}(z)\,z^\omega.
\eeq
We represent  the functions $B$ and $\tilde N$ through their inverse
Mellin transforms:
$$
B(Q,x)\,=\,\frac{1}{2\pi i}\int_C\,d\omega\,x^{-\omega}B(Q,\omega)\,;\,\,\,\,\,\,\,\,\,\,\,\,\,\,
\tilde N(2/Q,x)\,=\,\frac{1}{2\pi i}\int_C\,d\omega\,x^{-\omega}\tilde N(2/Q,\omega)\,.
$$
The equation  (\ref{DN1}) can be now expressed in the following form:
\beq
\label{DN2}
\frac{d B(Q,\omega)}{d(\ln Q^2)}\,=\,-\frac{\as\,N_c}{\pi}\left[\frac{Q^2}{4} \tilde N(2/Q,\omega)
\,+\, B(Q,\omega)\right]\,.
\eeq
The solution of the equation  (\ref{DN2}) is:
\beq
\label{solin}
B(Q,\omega)\,=\,(Q^2)^{-\bar\as}\left[-\,\frac{\bar\as}{4} \int^{Q^2}_{Q_0^2}
  \tilde N(2/Q^\prime,\omega)\,
(Q^{\prime 2})^{\bar\as}\,dQ^{\prime 2}\,+\,C(\omega)\right],
\eeq
with $\bar\as=\as\,N_c/\pi$. The function $C(\omega)$ 
should be determined by initial conditions at $Q=Q_0$.
 At very small $x$ and not too large $Q$ the 
function $N \simeq 1$ and  it is a slowly varying function of
$Q$. Hence,
 it can be taken
outside of the integral in (\ref{solin}) and we obtain the  solution
\beq
\label{sol}
 B(Q,\omega)\,\simeq\,-\frac{\bar\as/4}{1+\bar\as}\tilde N(2/Q,\omega)\,Q^2\,+\,
\frac{C(\omega)}{(Q^2)^{\bar\as}}\,.
\eeq
 As a result  for the correcting function $\Delta N$ we have:
\beq
\label{solDN}
\Delta N(2/Q,x)\,\simeq\,-\frac{\bar\as}{1+\bar\as}\tilde N(2/Q,x)\,+\,
\frac{4\,C(x)}{(Q^2)^{1+\bar\as}}\,.
\eeq

Recall that in the large $Q^2$ limit $r^2_\perp=4/Q^2$.
The  solution found in (\ref{solDN})
  is  in agreement with  assumptions made. 
The function $\Delta N$ decreases at small distances and is of
size  $O(\as)$
 compared to $\tilde N$. The argument presented above only aimed to
prove
the self consistency of our  approach. However, the complete numerical solution of the
equation (\ref{DN}) is still valid for phenomenological reasons.  The
initial conditions
for  the function $\Delta N$ should be chosen to fit  the experimental data.

\section{Numerical solution of the BFKL equation}

In this section we report on  the numerical solution of the BFKL equation, which is the linear
part of the equation (\ref{EQ}):
\begin{eqnarray}
\label{BFKLeq}
 N_{\rm BFKL}({\mathbf{x_{01}}},Y) \,=\,
 N_{\rm BFKL}({\mathbf{x_{01}}},Y_0)\, {\rm exp}\left[-\frac{2
\,C_F\,\as}{\pi} \,\ln\left( \frac{{\mathbf{x^2_{01}}}}{\rho^2}\right)(Y-Y_0)\right ]\,
+\\ \nonumber   \frac{C_F\,\as}{\pi^2} \times   \int_{Y_0}^Y dy \,  {\rm exp}\left[-\frac{2
\,C_F\,\as}{\pi} \,\ln\left( \frac{{\mathbf{x^2_{01}}}}{\rho^2}\right)(Y-y)\right ]\,
 \int_{\rho} \, d^2 {\mathbf{x_{2}}} 
\frac{{\mathbf{x^2_{01}}}}{{\mathbf{x^2_{02}}}\,
{\mathbf{x^2_{12}}}} \nonumber 
\,2\, N_{\rm BFKL}({\mathbf{x_{02}}},y)\,.
\end{eqnarray}
We restrict our analysis to $b=0$ case only. 
The initial conditions are assumed to be proportional to the gluon density  at $x_0=10^{-2}$:
\beq
\label{inibfkl}
N_{\rm BFKL}(\mathbf{x_{01}},x_0)\,=\, \frac{\as \pi  \mathbf{x_{01}^2}}{6 R^2}\,x
G(x_0,  4/\mathbf{x_{01}^2}).
\eeq

In order to obtain a solution of the equation (\ref{BFKLeq}) 
the  same method of iterations is applied. However,  for this case  
the convergence of the iterations is extremely  slow. Dozens of
iterations are required 
in order to get a final result. Fig. (\ref{figbfkl}) presents solutions of the BFKL
equation for $\as=0.25$ and for running $\as$ compared with the solutions of the equation
(\ref{EQ}) as well as of the DGLAP equation.

\begin{figure}[htbp]
\begin{tabular}{c c}
 \epsfig{file=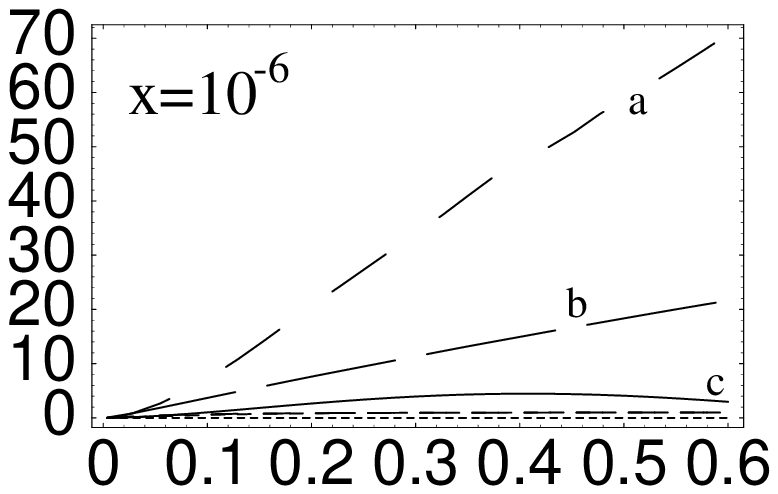,width=70mm, height=40mm}&
\epsfig{file=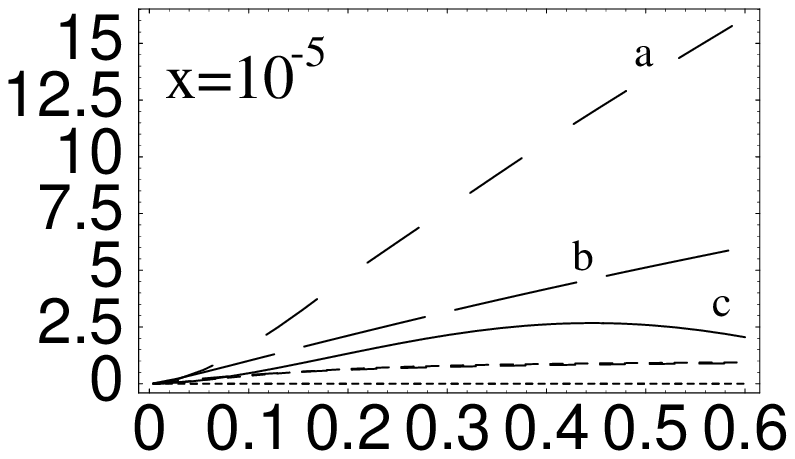,width=70mm, height=40mm}\\ 
 \epsfig{file=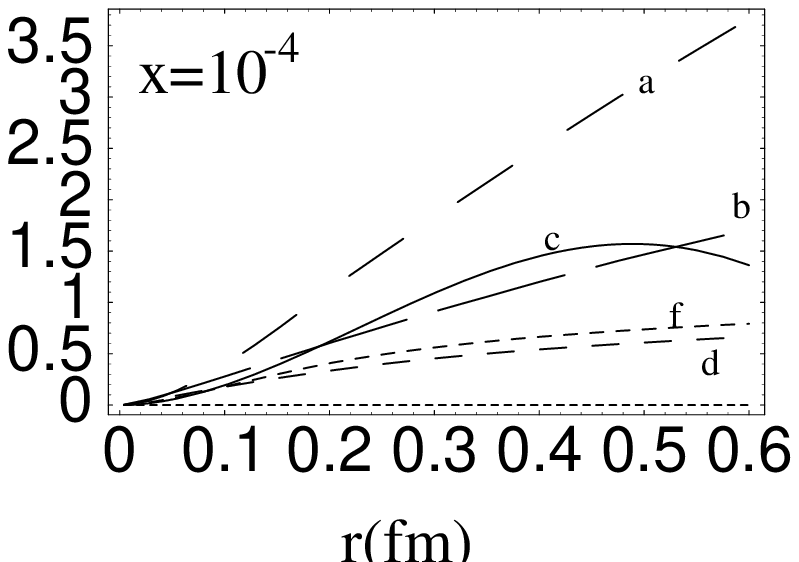,width=70mm, height=50mm}&
\epsfig{file=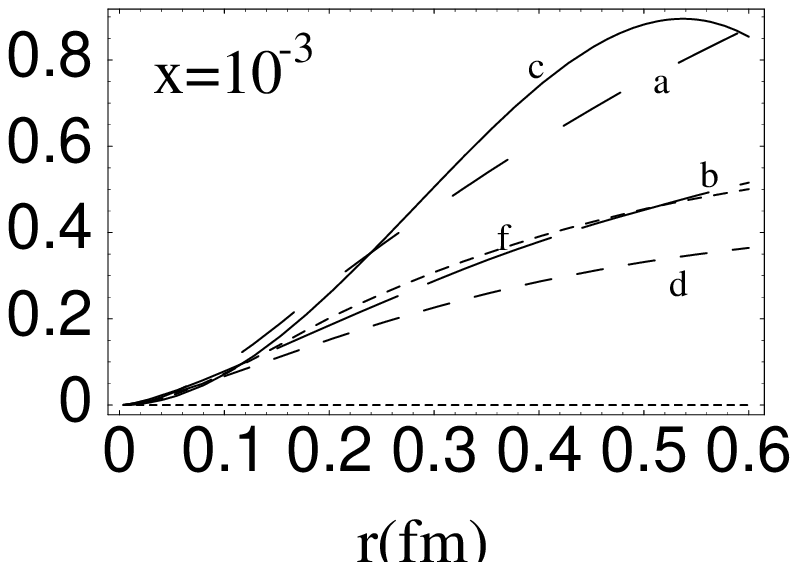,width=70mm, height=50mm}\\ 
\end{tabular}
  \caption[]{\it The function $N_{\rm BFKL}$ 
is plotted together with the  solution  of the equation (\ref{EQ}) 
 and  the DGLAP equation. a -  $N_{\rm BFKL}$ for running $\as$; 
b - $N_{\rm BFKL}$ for $\as=0.25$; c - DGLAP solution in GRV parameterization; 
d - $\tilde N$ for   $\as=0.25$; f - $\tilde N$ for  running $\as$. }
\label{figbfkl}
\end{figure}

It is seen from the Fig.  (\ref{figbfkl}) that the BFKL equation predicts  a very steep growth
of the cross section as $x$ decreases. The function $N_{\rm BFKL}$ rapidly diverges 
from the solutions of the equation (\ref{EQ}) and of the DGLAP equation. 
This numerical observation
is  in fair agreement with previous theoretical considerations about
the dynamics of the BFKL 
equation. The solution of the BFKL equation touches the unitarity bound.

However, the results of BFKL dynamics is unlikely ever to be
seen. The nonlinear effects accounted for
in the equation (\ref{EQ}) set in at  very small distances thus suppressing the BFKL
divergence. 

\section{Summary}

In  the present work we suggested a new approach to DIS based on first summing
 all twist contributions in the leading $\ln x$ approximation. Then, we
wrote down
a linear evolution equation for the correcting function which incorporates the correct DGLAP
kernel in the leading $\ln Q^2$ approximation.

The method of iterations was successfully applied to the solution of the
nonlinear 
integral equation (\ref{EQ}). A key observation is  the nonlinearity of the equation,
and that  a good choice for the zero iteration, insures a rapid
 convergence.

The BFKL equation is solved numerically. 
Its solution blows up for  low values of $x$, and
this is the origin for the possible violation of  unitarity. However, the nonlinear effects
suppress the growth of the cross section. These effects set in at such small distances that we
do not actually  ever expect to  see the BFKL dynamics.

The nonlinear BFKL-type equation (\ref{EQ}) is solved numerically. The solution predicts 
a saturation of the function $\tilde N$ at large distances. The 
 dipole cross section obtained 
is not saturated as a function of
 $x$.  Due to $b$ integration   it grows logarithmically  
 with decrease of $x$. 
Fig. (6) shows that the solution to the non-linear evolution equation generates a dipole cross
 section which is quite different from the predictions of the
 Golec-Biernat and  Wusthoff model 
\cite{WG}.
 This model is widely used
 for estimating  saturation effects at HERA, and our calculations
demonstrate that we 
should be very careful with
 the conclusions based on such estimates especially at lower $x$.

We demonstrated that the solution to the non-linear equation is quite 
different from the models that have been used
for estimates of the saturation effects \cite{rep,WG} (see Fig. (2)). 
However, all these models can be used as a first
 iteration of the non-linear equation which leads to faster convergence of 
the numerical procedure.

We found  that the non-linear effects are very important for the extrapolations of improved
 knowledge on the parton
 distributions to higher energies such as the THERA and LHC energies. 
Fig. (7) shows that damping due to non-linear
 effects leads to suppression of a factor of $2 \div 3$
  in comparison with the linear DGLAP evolution.

The  solution found for the nonlinear equation was used  to
estimate the saturation scale
$Q_s(x)$.   In spite of considerable uncertainty in the value of the
 saturation scale  we predict that it grows with decreasing $x$  
starting from about
 1(GeV) at
$x=10^{-3}$  and reaching around 10(GeV) at $x=10^{-7}$, in accordance
with the theoretical 
expectations
 \cite{GLR,MUQI,MV,SAT}.

We proved that the correcting function $\Delta N$ is concentrated at the moderate values of 
$x$. This fact
 demonstrates
 the self-consistency of our approach, as $\Delta N$ cannot be large and
can be treated using 
the linear DGLAP-type of
 equations. The full phenomenological study of the value of $\Delta N$   
and its $x$ and $Q^2$ dependence 
will be published  elsewhere.

We hope that our approach will be useful for the extrapolation of the HERA 
parton distributions to higher energies 
(lower $x$). Of course much further work will be required before this
area of QCD is fully understood.

{\bf Acnowledgements:}
The authors are very much indebted to our coauthors and friends with
whom we discussed our approach on  a everyday basis Ian Balitsky, Jochen
Bartels ,
Krystoff Golec Biernat, Larry
McLerran, Dima Kharzeev, Yuri Kovchegov and  Al Mueller for their help and
fruitful discussions on the subject. E.G. ,  E. L.  thank BNL
Nuclear
Theory Group and  DESY Theory group
for their hospitality and  
creative atmosphere during several stages of this work.

This research was supported in part by the BSF grant $\#$
9800276, by the GIF grant $\#$ I-620-22.14/1999
  and by
Israeli Science Foundation, founded by the Israeli Academy of Science
and Humanities.


\begin{thebibliography}{99}


\bibitem{DGLAP} V. N. Gribov and L. N. Lipatov, {\it Sov. J. Nucl. Phys} {\bf 15} (1972)
                438; G. Altarelli and G. Parisi, {\it Nucl. Phys.}
                {\bf B 126} (1977) 298; Yu. l. Dokshitser, {\it Sov. Phys. JETP} {\bf 46}
                (1977) 641.

\bibitem{GLR} L. V. Gribov, E. M. Levin, and M. G. Ryskin, {\it Nucl. Phys.} {\bf B 188}
(1981) 555.


\bibitem{HT}
J. Bartels, \plb{ 298}{93}{204},\zpc{ 60}{93}{471};\\
E.M.  Levin, M.G. Ryskin, and A.G. Shuvaev, \npb{ 387}{92}{589}.
                                                                                
\bibitem{KKM}
M. A.  Kimber, J. Kwiecinski, and A. D. Martin,  IPPP-01/01, DCPT/01/02,  hep-ph/0101099.

\bibitem{THERA}

M. Klein, {\it ``THERA-electron proton scattering at $\sqrt{s} \approx
  1\,TeV$''}, talk given at DIS'2000, Liverpool, April 25 - 30, 2000.

\bibitem{MUQI}
A. H. Mueller and J. Qiu, {\it Nucl. Phys.} {\bf B 268} (1986) 427.

\bibitem{MV}
L. McLerran and R. Venugopalan,{\it Phys. Rev. } {\bf D 49} (1994)
2233, 3352; {\bf D 50} (1994) 2225, {\bf D 53} (1996) 458, {\bf D 59} (1999)
094002.   
\bibitem{SAT}
E.   Levin and M.G. Ryskin, {\it Phys. Rep.} {\bf 189} (1990) 267;\\
J. C. Collins and J. Kwiecinski, \npb{ 335}{90}{89};\\
J. Bartels, J. Blumlein, and G. Shuler, \zpc{ 50}{91}{91};\\
E. Laenen and E. Levin, \arnps{44}{94}{199}
and references therein;\\
A. L. Ayala, M. B. Gay Ducati,  and E. M. Levin, \npb{ 493}{97}{305},
{\bf B 510} (1990) 355;\\  Yu. Kovchegov, \prd{ 54}{96}{5463}, {\bf D 55} (1997) 5445,
{\bf D 61} (2000) 074018;\\ A. H. Mueller,
{\it Nucl. Phys.} {\bf B 572} (2000) 227,
{\bf B 558} (1999) 285;\\ Yu. V. Kovchegov, A. H. Mueller,
\npb{ 529}{98}{451}.

\bibitem{ELTHEORY}
J. Jalilian-Marian, A. Kovner, L. McLerran,  and  H.
Weigert, \prd{ 55}{97}{5414};
J. Jalil H. Weigert, NORDITA-2000-34-HE,  hep-ph/0004044;
J. Jalilian-Marian, A. Kovner, and  H.
Weigert, \prd{ 59}{99}{014015};
J. Jalilian-Marian, A. Kovner, A.
Leonidov, and  H. Weigert, \prd{ 59}{99}{034007},
Erratum-ibid. \prd{ 59}{99}{099903};
A. Kovner, J.Guilherme Milhano, and  H. Weigert,
OUTP-00-10P, NORDITA-2000-14-HE,  hep-ph/0004014;
 H. Weigert, NORDITA-2000-34-HE,  hep-ph/0004044.
  
 
 \bibitem{BA}
Ia. Balitsky, {\it Nucl.Phys. } {\bf B 463}  (1996) 99.

\bibitem{KO}
Yu. Kovchegov,
{\it Phys. Rev.} {\bf D 60} (2000) 034008. 
                                                                                                                         
\bibitem{ILM}
E. Iancu, A. Leonidov, and L. McLerran, {``Nonlinear Gluon Evolution in the
Color Glass Condensate"}, BNL-NT-00/24,  hep-ph/0011241.  

\bibitem{MU94}
A. H.  Mueller, {\it  Nucl. Phys.}  {\bf B 415} (1994) 373.

\bibitem{DOF3}
A. H. Mueller, {\it Nucl. Phys.} {\bf B 335} (1990) 115.

\bibitem{WF}
N. N. Nikolaev and B. G. Zakharov, {\it Z. Phys.} {\bf C 49} (1991) 607;
E. M. Levin, A. D. Martin, M. G.  Ryskin, and T. Teubner, {\it Z. Phys.} {\bf
C 74} (1997) 671.


\bibitem{GRIB}
V. N. Gribov, {\it Sov. Phys. JETP} {\bf 30} (1970) 709.   

\bibitem{BFKL} E. A. Kuraev, L. N. Lipatov, and F. S. Fadin, {\it Sov. Phys. JETP}
                {\bf 45} (1977) 199; Ya. Ya. Balitsky and L. N. Lipatov,
                {\it Sov. J. Nucl. Phys.} {\bf 28} (1978) 22 .

\bibitem{HTM}
J. Bartels, K. Golec-Biernat and K.  Peters, {\it  Eur. Phys. J.} {\bf C17}
(2000) 121,  hep-ph/0003042;  
 E. Gotsman, E. Levin, L. McLerran, and  K. Tuchin, 
{\it Nucl. Phys.} {\bf B}, in  press,  hep-ph/0008280.
  
\bibitem{DL}
A. Donnachie and P. V. Landshoff, \npb{ 244}{84}{322},{\bf B 267} (1986) 690;
\plb{ 296}{92}{227};\zpc{ 61}{94}{139}.
 

\bibitem{DOF1}
A. Zamolodchikov, B. Kopeliovich, and L. Lapidus, {\it JETP Lett.} {\bf 33}
(1981) 595.

\bibitem{DOF2}
E. M.  Levin and M. G.  Ryskin, {\it Sov. J. Nucl. Phys.} {\bf 45} (1987) 150.

\bibitem{rep}
E. Gotsman, E. Levin, and  U. Maor, {\it Nucl. Phys.} {\bf B 464} (1996) 251;
{\bf B 493} (1997) 354;\\
E. Gotsman, E. Levin, M. Lublinsky, U. Maor, E. Naftali, and  K. Tuchin,   hep-ph/0010198,
DESY-00-149.

\bibitem{me}
E. Gotsman, E. Levin, M. Lublinsky, U. Maor, and  K. Tuchin,   hep-ph/0007261.


\bibitem{LT} 
Yu. Kovchegov,  { \it Phys. Rev.} {\bf D 61} (2000) 074018;
 E. Levin and K. Tuchin, {\it Nucl. Phys.} {\bf B 573} (2000) 833; hep-ph/0012167.

\bibitem{Braun} M. Braun, {\it Eur. Phys. J.} {\bf C 16} (2000) 337; hep-ph/0101070.


\bibitem{KMS} J Kwiecinski, A. D. Martin, and A. M. Stasto, {\it Phys. Rev.}
 {\bf D 56} (1997) 3991.

\bibitem{GLMSOFT}
E. Gotsman, E. Levin,  and U. Maor, \plb{ 452}{99}{287}; \prd{ 49}{94}{4321};
\plb{ 304}{93}{199},\zpc{ 57}{93}{672}.

\bibitem{HERAPSI}
H1 Collaboration: S. Aid et al., \npb{ 472}{96}{3};\\
ZEUS Collaboration: M. Derrick et al., \plb{ 350}{96}{120}.

\bibitem{GRV}
M. Gluck, E. Reya, and A. Vogt, {\it Eur. Phys. J.} {\bf C 5} (1998) 461.


\bibitem{WG}
K. Golec-Biernat and M. W\"{u}sthoff, \prd{ 59}{99}{014017}.

\bibitem{EKL} R. K. Ellis, Z. Kunszt, and E. M. Levin, {\it Nucl. Phys.} {\bf B 420} (1994) 517. 



\end{thebibliography}
\end{document}